\documentclass{kapproc}

\normallatexbib

\begin{document}
\input epsf 

\def\pp{{\, \mid \hskip -1.5mm =}}
\def\cL{{\cal L}}
\def\be{\begin{equation}}
\def\ee{\end{equation}}
\def\bea{\begin{eqnarray}}
\def\eea{\end{eqnarray}}
\def\tr{{\rm tr}\, }
\def\nn{\nonumber \\}
\def\e{{\rm e}}
\def\D{{D \hskip -3mm /\,}}

\def\SEH{S_{\rm EH}}
\def\SGH{S_{\rm GH}}
\def\AdS5{{{\rm AdS}_5}}
\def\S4{{{\rm S}_4}}
\def\gfv{{g_{(5)}}}
\def\gfr{{g_{(4)}}}
\def\SC{{S_{\rm C}}}
\def\RH{{R_{\rm H}}}

\def\wlBox{\mbox{
\raisebox{0.1cm}{$\widetilde{\mbox{\raisebox{-0.1cm}\fbox{\ }}}$}}}
\def\htBox{\mbox{
\raisebox{0.1cm}{$\hat{\mbox{\raisebox{-0.1cm}{$\Box$}}}$}}}

\def\K{\left(k - (d-2) \e^{-2\lambda}\right)}

\articletitle{The  
de Sitter/Anti- de Sitter Black 
Holes phase transition?}

\author{Shin'ichi NOJIRI}
\affil{Department of Applied Physics \\
National Defence Academy,
Hashirimizu Yokosuka 239-8686, JAPAN}
\email{nojiri@cc.nda.ac.jp, snojiri@yukawa.kyoto-u.ac.jp}

\author{Sergei D. Odintsov}
\affil{Tomsk State Pedagogical University \\
Tomsk, RUSSIA and
Instituto de Fisica de la Universidad de Guanajuato, \\
Lomas del Bosque 103, Apdo. Postal E-143, 
37150 Leon,Gto., MEXICO }
\email{odintsov@ifug5.ugto.mx, odintsov@mail.tomsknet.ru}

\chaptitlerunninghead{The  
de Sitter/Anti- de Sitter Black 
Holes phase transition?}

\begin{abstract}

We investigate the  Schwarzschild-Anti-deSitter (SAdS) and SdS BH 
thermodynamics in 5d higher derivative
gravity. 
The interesting feature of higher derivative gravity is 
the possibility for negative (or zero) SdS (or SAdS) BH 
entropy which  depends on the parameters of higher derivative 
terms. The appearence of negative entropy 
may indicate a new type instability where a transition between
SdS (SAdS) BH with negative entropy to SAdS (SdS) BH with positive 
entropy  would occur or where definition of entropy should be modified.
\end{abstract}


BH thermodynamics is quite attractive, as it provides
the understanding of gravitational physics at extremal conditions.
Moreover, it has been realized recently that AdS BH may be relevant 
in the study of AdS/CFT correspondence. Hence, there appears nice way to 
decribe strong coupling gauge theories via their gravitational duals.
In the present work we discuss the thermodynamics of dS and AdS BHs
in higher derivative gravity. The fundamental issue of entropy
for such objects leads to some interesting conclusions.

We start from the following action of $d$ dimensional 
$R^2$-gravity with cosmological constant. 
The action is given by:
\be
\label{vi}
S=\int d^{d+1} x \sqrt{-g}\left\{a R^2 +b R_{\mu\nu} R^{\mu\nu}
+ c R_{\mu\nu\xi\sigma} R^{\mu\nu\xi\sigma}
+ {1 \over \kappa^2} R - \Lambda 
\right\} \ .
\ee
We discuss the relation between SdS and SAdS BHs 
based on entropy considerations.
For simplicity, we consider $c=0$ 
case in (\ref{vi}) for most results.
When $c=0$, Schwarzschild-anti de Sitter space is 
an exact solution:
\bea
\label{SAdSA}
&& ds^2 = - \e^{2\nu (r)} dt^2 + \e^{-2\nu (r)} dr^2 
+ r^2 \sum_{i,j=1}^{d-1} \tilde g_{ij} dx^i dx^j\ ,\nn
&& \e^{2\nu}=\e^{2\nu_0}\equiv 
{1 \over r^{d-2}}\left(-\mu + {kr^{d-2} \over d-2} 
+ {r^d \over l^2}\right)\ .
\eea
Here $\tilde g_{ij}$ is the metric of the Einstein manifold, 
which is defined by $\tilde R_{ij}=k g_{ij}$.  
$\tilde R_{ij}$ is the Ricci curvature given by $\tilde g_{ij}$ 
and $k$ is a constant. For example, one has $k=d-2$ for 
$d-1$-dimensional unit sphere, $k=-(d-2)$ for $d-1$-dimensional 
unit hyperboloid, and $k=0$ for flat surface. 
The curvatures have the following form:
$\hat R=-{d(d+1) \over l^2}$ and 
$\hat R_{\mu\nu}= - {d \over l^2}\hat G_{\mu\nu}$. 
In (\ref{SAdSA}), $\mu$ is the parameter corresponding to mass 
and the scale parameter $l$ is given by solving the following 
equation:
\be
\label{llA}
0={d^2(d+1)(d-3) a \over l^4} + {d^2(d-3) b \over l^4} \nn
- {d(d-1) \over \kappa^2 l^2}-\Lambda\ .
\ee
When the solution for $l^2$ (\ref{llA}) is positive 
(negative), the spacetime is asymptotically anti-de Sitter 
(de Sitter) space and especially if $\mu=0$, the solution 
expresses the pure anti-de Sitter or de Sitter space.  
If $\Lambda=0$, there is asymptotically flat (Minkowski) 
solution, where ${1 \over l^2}=0$.  
In the following we concentrate on the case of $d=4$. 

The calculation of thermodynamical quantites like free 
energy $F$, the entropy ${\cal S}$ and the energy $E$ may be done 
with the help of the following method: After 
Wick-rotating the time variable by $t \to i\tau$, the
free energy $F$ can be obtained from the action $S$ (\ref{vi})
where the classical solution is substituted : $F=-TS$. 
Substituting eq.(\ref{llA}) into (\ref{vi}) in the case of $d=4$ 
with $c=0$, one gets
\bea
\label{sr}
S &=& - \int d^{5}x \sqrt{-G} \left( {8 \over l^2\kappa^2}
 - {320 a \over l^4} -{64 b \over l^4} \right) \nn
&=& -{V_{3} \over T}\int ^{\infty}_{r_{H}} dr r^{3}
\left( {8 \over l^2\kappa^2}- {320 a \over l^4}
 -{64 b \over l^4} \right)
\eea
Here $V_{3}$ is the volume of 3d sphere and we assume $\tau$
has a period of ${1\over T}$.  The expression of $S$ contains
the divergence coming from large $r$. In order to subtract the
divergence, we regularize $S$ (\ref{sr}) by cutting off the
integral at a large radius $r_m$ and subtracting the
solution with $\mu =0$:
\be
\label{regaction}
S_{\rm reg}=
-{V_{3} \over T}\left\{ \int ^{r_m}_{r_{H}} dr r^{3}
-{\e^{\nu(r=r_m)} \over \e^{\nu(r=r_m;\mu =0)} }
\int ^{r_m}_{0} dr r^{3} \right\} 
\left( {8 \over \kappa^2 l^{2} }- {320 a \over l^{4}}
 -{64 b \over l^{4}} \right)
\ee
The factor $\e^{\nu(r=r_m)}/\e^{\nu(r=r_m;\mu =0)}$ is chosen
so that the proper length of the circle which corresponds to the
period ${1 \over T}$ in the Euclidean time at $r=r_m$
concides with each other in the two solutions.  Taking 
$r_m \to \infty$, one finds, as found in \cite{NOOr2a}, 
\bea
F= -V_{3}\left( {l^2 \mu \over 8} 
 - {r_H^4  \over 4}\right) \left( {8 \over l^2\kappa^2}
 - {320 a \over l^4} -{64 b \over l^4} \right)
\eea
The horizon radius $r_{h}$ is given by solving 
the equation $e^{2\nu_0(r_H)}=0$ in (\ref{SAdSA}):
\bea
\label{rh1}
r_{H}^{2}=-{k l^{2} \over 4} + {1\over 2}
\sqrt{{k^2 \over 4}l^{4}+ 4\mu l^{2} } \; .
\eea
The Hawking temperature $T_H$ is
\be
\label{ht1}
T_H = {k \over 4\pi r_{H}} +{r_{H} \over \pi l^2} 
\ee
where $'$ denotes the derivative with respect to $r$. From 
the above equation (\ref{ht1}), $r_{H}$ can be rewritten  in terms of 
$T_{H}$ as
\bea
\label{rHTH}
r_{H}={1\over 2}\left( \pi l^{2} T_{H} \pm
\sqrt{(\pi l^{2} T_{H})^{2} -kl^{2}} \right)
\eea
In (\ref{rHTH}), the plus sign corresponds to $k=-2$ or $k=0$ 
case and the minus sign to $k=2$ case.\footnote{
When $k=2$, as we can see from (\ref{rh1}) and (\ref{ht1}), $r_H$, 
and also $T_H$, are finite in the limit of $l\rightarrow \infty$, 
which corresponds to the flat background. Therefore we need to 
choose the minus sign in (\ref{rHTH}) for $k=2$ case.}
One can also rewrite $\mu $ by using $r_{H}$ as 
$\mu ={r_{H}^{4} \over l^{2}}+{kr_{H}^{2} \over 2}$. 
Then we can rewrite $F$ using $r_{H}$ as
\be
\label{FF}
F= -{V_{3} \over 8}r_{H}^{2} \left( {r_{H}^{2} \over l^{2}}
 - {k \over 2} \right)\left( {8 \over \kappa^2} 
 - {320 a \over l^2} -{64 b \over l^2} \right) \; .
\ee
Then the entropy ${\cal S }= -{dF \over dT_H}$ and the 
thermodynamical energy $E = F+T{\cal S}$ can be obtained 
as follows \cite{NOOr2a}:
\be
\label{entA}
{\cal S }=4V_{3}\pi r_H^3 
\left( {1 \over \kappa^2}- {40 a \over l^2}
 -{8 b \over l^2} \right)\ ,\quad 
%\label{enerA}
E= 3V_{3}\mu 
\left( {1 \over \kappa^2}- {40 a \over l^2}
 -{8 b \over l^2} \right)\ , 
\ee
This seems to indicate that the contribution from 
the $R^2$-terms can be absorbed into the redefinition:
\be
\label{B16A}
{1 \over \tilde\kappa^2}={1 \over \kappa^2} - {40a \over l^2} 
 - {8b \over l^2}\ ,
\ee
although this is not true for $c\neq 0$ case. 

On the other hand, by using the surface counter term 
method \cite{CNO} , 
one gets the following expression for the conserved mass $M$:
\be
\label{ABC11}
M={3l^2 \over 16}V_3 \left({1 \over \kappa^2} - {40a \over l^2}
 - {8b \over l^2} - {4c \over l^2}\right)\left(k^2  
+ {16\mu \over l^2}\right)\ .
\ee
One can also start from the expression for $M$ with $c=0$ as 
the thermodynamical energy $E$:
\be
\label{ABCE1}
E=3V_3 \left({1 \over \kappa^2} - {40a \over l^2}
 - {8b \over l^2} \right)\left({k^2l^2 \over 16} 
+ {\mu \over l^2}\right)
\ee
The expression of energy $E$ (\ref{ABCE1}) is different from that 
in (\ref{entA}) by a first $\mu$-independent term, which  comes 
from the AdS background. 
Using the thermodynamical relation $d{\cal S}={dE \over T}$, 
we find 
\be
\label{ABGE3}
{\cal S}=\int {dE \over T_H} = \int dr_H {dE \over d\mu}
{d \mu \over d r_H}{1 \over T_H} = {V_{3}\pi r_H^3 \over 2}
\left( {8 \over \kappa^2}- {320 a \over l^2}
 -{64 b \over l^2} \right) + {\cal S}_0\ .
\ee
Here $S_0$ is a constant of the integration. Up to the constant 
$S_0$, the expression (\ref{ABGE3}) is identical with (\ref{entA}). 
We should note that the entropy ${\cal S}$ (\ref{entA}) 
becomes negative, when 
\be
\label{EnS1}
{8 \over \kappa^2}- {320 a \over l^2}
 -{64 b \over l^2}<0.\ .
\ee
This is true even for the expression (\ref{ABGE3}) for the 
black hole with large radius $r_H$ since $S_0$ can be neglected 
for the large $r_H$. 

We now investigate in more detail what happens when 
Eq.(\ref{EnS1}) is satisfied. First we should note $l^2$ is 
determined by (\ref{llA}), which has, in case of $d=4$, the 
following form:
\be
\label{llA4}
0={80 a + 16 b \over l^4} - {12 \over \kappa^2 l^2} - \Lambda\ ,
\ee
There are two real solutions for $l^2$ when 
${6 \over \kappa^2} + \left(80a + 16b\right)\Lambda \geq 0$ 
and the solutions are given by
\be
\label{lll2}
{1 \over l^2}={{6 \over \kappa^2}\pm \sqrt{
{6 \over \kappa^2} + \left(80a + 16b\right)\Lambda} 
\over 80a + 16b}\ .
\ee
Suppose $\kappa^2>0$. Then if 
\be
\label{lll3}
\left(80a + 16b\right)\Lambda>0\ ,
\ee
one solution is positive but another is negative. 
Therefore there are both of the asymptotically AdS solution 
and asymptotically dS one. Let us denote the positive 
solution for $l^2$ by 
$l_{\rm AdS}^2$ and the negative one by $-l_{\rm dS}^2$:
\be
\label{lll4}
l^2=l_{\rm AdS}^2,\ -l_{\rm dS}^2\ ,\quad 
l_{\rm AdS}^2,\ l_{\rm dS}^2 >0\ .
\ee
Then when the asymptotically AdS solution is chosen, the entropy 
(\ref{ABGE3}) has the following form: 
\be
\label{lll5}
{\cal S}_{\rm AdS}= {V_{3}\pi r_H^3 \over 2}
\left( {8 \over \kappa^2}- {320 a + 64b \over l_{\rm AdS}^2}
\right) \ .
\ee
Here we have chosen ${\cal S}_0=0$. On the other hand, 
when the solution is asymptotically dS, the entropy 
(\ref{ABGE3}) has the following form: 
\be
\label{lll6}
{\cal S}_{\rm dS}= {V_{3}\pi r_H^3 \over 2}
\left( {8 \over \kappa^2} + {320 a + 64b \over l_{\rm dS}^2}
\right) \ .
\ee
When
\be
\label{lll7}
{8 \over \kappa^2}- {320 a + 64b \over l_{\rm AdS}^2}<0\ ,
\ee
the entropy ${\cal S}_{\rm AdS}$ (\ref{lll5}) is negative!

There are different points of view to this situation. Naively, 
one can assume that above condition is just the equation to
remove the non-physical domain of theory parameters. However, it 
is difficult to justify such proposal.
Why for classical action on some specific background 
there are parameters values which are not permitted?
Moreover, the string/M-theory and its compactification 
would tell us what are the values of the theory parameters.

 From another side, one can conjecture that classical 
thermodynamics is not applied here and negative entropy simply 
indicates to new type of instability in asymptotically AdS black 
hole physics. Indeed, when Eq.(\ref{lll7}) is satisfied, since 
$80a + 16b >0$ (same range of parameters!), the entropy 
${\cal S}_{\rm dS}$ (\ref{lll6}) for asymptotically dS solution 
is positive. In other words, may be the asymptotically dS solution 
would be preferrable?

On the other hand, when 
\be
\label{lll8}
{8 \over \kappa^2}+ {320 a + 64b \over l_{\rm dS}^2}<0\ ,
\ee
the entropy ${\cal S}_{\rm dS}$ in (\ref{lll6}) is negative and the 
asymptotically dS solution is instable (or does not exist).
 In this case, since $80a + 16b <0$, the entropy 
${\cal S}_{\rm AdS}$ in (\ref{lll5}) for asymptotically 
AdS solution is positive and the asymptotically AdS solution 
would be preferrable. Expression for the AdS black hole mass in 
(\ref{ABCE1}) tells that when 
${8 \over \kappa^2}- {320 a + 64b \over l_{\rm AdS}^2}=0$, the 
AdS black hole becomes massless then there would occur the 
condensation of the black holes, which would make the transition to 
the dS black hole. On the other hand, when ${8 \over \kappa^2} + 
{320 a + 64b \over l_{\rm dS}^2}=0$, the dS black hole becomes 
massless then there would occur the condensation of the black 
holes and the AdS black hole would be produced. 
Note that above state with zero entropy (and also zero free 
energy and zero conserved BH mass) is very interesting. Perhaps, 
this is some new state of BHs. As we saw that is this state which 
defines the border between physical SAdS (SdS) BH with positive 
entropy and  SdS (SAdS) BH with negative entropy.

Hence, there appeared some indication that some new type of 
phase transition (or phase transmutation) between
SdS and SAdS BHs in higher derivative gravity occurs.
Unfortunately, we cannot suggest any dynamical formulation to 
describe explicitly such phase transition (it is definitely 
phase transition not in standard thermodynamic sense).

Let us consider now the entropy for Gauss-Bonnet case, 
where $a=c$ and $b=-4c$ in (\ref{vi}). For this 
purpose, we use the thermodynamical relation $d{\cal S}={dE \over T}$. For 
the Gauss-Bonnet case, the energy (\ref{ABC11}) has the 
following form \cite{CNO}:
\be
\label{EnS2}
E=M={3l^2 \over 16}V_3 \left({1 \over \kappa^2} - {12c \over l^2}
\right)\left(k^2  + {16\mu \over l^2}\right)
\ee
We also found \cite{CNO}
\be
\label{EnS3}
\mu = {1 \over 2l^2}\left(k\epsilon - {1 \over 2}\right)^{-1}
\left\{ ( 2\epsilon - 1) r_H^4 
 - kr_H^2 l^2 \right\}\ .
\ee
Here $\epsilon\equiv {c\kappa^2 \over l^2}$. 
Then using (\ref{EnS2}), (\ref{EnS3}), and the expression 
of the Hawking temperature, 
\be
\label{GBxxx}
4\pi T_H = {1 \over 2}\left(ck + {r_H^2 \over 2\kappa^2 }
\right)^{-1}\left[ {kr_H \over \kappa^2 } 
 - {8c r_H^3 \over l^4}
 + {4 r_H^3 \over \kappa^2 l^2} - {2 Q^2 \over 3 g^2 r_H^3} 
\right]\ ,
\ee
the entropy can be 
obtained as
\bea
\label{EnS4} 
\lefteqn{{\cal S}=\int {dE \over T_H} = \int dr_H {dE \over d\mu}
{\partial \mu \over \partial r_H} {1 \over T_H} } \\
&& =  {V_3 \over \kappa^2}\left({1 - 12 \epsilon 
\over 1-4\epsilon} \right)\left(4\pi r_H^3 
+ 24 \epsilon k \pi r_H\right) + {\cal S}_0\ .
\eea
Here ${\cal S}_0$ is a constant of the integration, which 
could be chosen to be zero if we assume ${\cal S}=0$ when 
$r_H=0$.  
When $\epsilon=0$ ($c=0$), the expression reproduces the 
standard result 
\be
\label{EnS5} 
{\cal S}\rightarrow {4\pi V_3 r_H^3 \over \kappa^2}\ .
\ee
The entropy (\ref{EnS4}) becomes negative (at least for the 
large black hole even if ${\cal S}_0\neq 0$) when 
\be
\label{EnS6}
{1 \over 12}<\epsilon<{1 \over 4}\ .
\ee
Therefore the unitarity might be broken in this region but 
it might be recovered when $\epsilon>{1 \over 4}$. Even in case 
$\epsilon<0$ ($k=2$), the entropy becomes negative when
\be
\label{EnSS1}
r_H^2 < - 12 \epsilon \ ,
\ee
if ${\cal S}_0=0$. Then the small black hole might be 
unphysical. 

The fact discovered here-that entropy for S(A)dS BHS in 
gravity with higher derivatives terms may be easily done to 
be negative by the corresponding choice of parameters is quite 
remarkable. It is likely that thermodynamics for black holes 
with negative entropies should be reconsidered. In this respect 
one possibility would be to redefine the gravitational entropy for higher 
derivative gravity.

 \begin{acknowledgments}
The authors are grateful to M. Cveti\v c for collaboration. SDO would 
like to thank the organizers of First Mexican Meeting on Math. and 
Exp. Physics for 
hospitality. 
The work by SN is supported in part by the Ministry of Education, 
Science, Sports and Culture of Japan under the grant n. 13135208.
\end{acknowledgments}

%\begin{thebibliography}{1}
%\begin{thebibliography}{99}

\end{document}